\documentclass{iucr}              
\journalcode{J}

\usepackage{amsmath}
\usepackage{mathrsfs}
\usepackage{epsfig}
\usepackage{graphicx}
\usepackage{fancybox}

\usepackage{cases,multirow}
\usepackage{url,xcolor}
\usepackage[percent]{overpic}

\usepackage{enumitem}

\usepackage{comment}
\usepackage{bm}

\newtheorem{thm}{\bf Theorem}

\newtheorem{rem}{ Remark}[section]
\graphicspath{{images/}}
\newtheorem{prop}{\bf Proposition}
\newtheorem{coro}{\bf Corollary}[section]

\newtheorem{assump}{\bf Assumption}
\newtheorem{definit}{\bf Definition}

\begin{document}

\title{Analyzer Free Linear Dichroic Ptychography}

\ifx \aff \undefined
\author{Huibin Chang$^1$ and Matthew A Marcus$^2$ and Stefano Marchesini$^{2,3}$ }

 \date{%
    $^1$School of Mathematical Sciences, Tianjin Normal University, Tianjin, China\\%
    $^2$ Advanced Light Source, Lawrence Berkeley National Laboratory, Berkeley, CA, USA\\[2ex]
    $^3$ Computational Research Division, Lawrence Berkeley National Laboratory, Berkeley, CA, USA
    \today
}

\else
\author[a*]{Huibin}{ Chang}
\author[b]{Matthew A}{ Marcus}
\cauthor[c]{Stefano}{ Marchesini }{changhuibin@gmail.com, smarchesini@lbl.gov}

\aff[a]{School of Mathematical Sciences, Tianjin Normal University, Tianjin, \country{China}} 
\aff[b]{Advanced Light Source, Lawrence Berkeley National Laboratory, Berkeley, CA, \country{USA}}
\aff[c]{Computational Research Division, Lawrence Berkeley National Laboratory, Berkeley, CA, \country{USA}}

\fi


\maketitle
\begin{abstract}
Linear-dichroism is an important tool to characterize the transmission matrix and determine the crystal or orbital orientation in a material. In order to gain  high resolution mapping of the transmission properties of such materials, we introduce the linear-dichroism scattering model in ptychographic imaging, and then develop an efficient two-stage reconstruction algorithm. Using proposed algorithm,   the  dichroic transmission matrix without an analyzer can be recovered by using ptychography measurements with as few as three different polarization  angles, with the help of an empty region to remove phase ambiguities.  
\end{abstract}

\section{Introduction}

Ptychography is an imaging method which offers spatial resolution beyond the limitations of focusing optics \cite{rodenburg2008ptychography}.  This is done by collecting far-field diffraction patterns produced by the sample upon being illuminated with probe beams at an array of spots such that the probes overlap.
By enforcing consistency of the reconstructed sample in the overlap regions,  the transmission image of the sample can be reconstructed without knowledge of the phase of the diffraction patterns.  By processing the set of diffraction images through a reconstruction algorithm, it is also  possible to reconstruct the probe beam in amplitude and phase as well as the object's transmission at a set of points whose spacing is less than the size of the probe beam.  The reconstruction algorithm is thus the heart of the ptychographic method.  Because ptychography offers spatial resolution which is not limited by that of any focusing optic, it is particularly suited to X-ray microscopy because optics with large numerical aperture, hence small spots, do not exist.

In most implementations of ptychography, it is assumed that the sample is isotropic in that the transmitted electric field at any point is a (complex) scalar times the incident field.  This is a good approximation for most cases, in visible-light and X-ray microscopy.  However, many materials have orientational order on length scales exceeding the probe size.  Such materials can be locally birefringent and/or dichroic.  In the X-ray region, this effect occurs when the incident energy is resonant with transitions to unoccupied orbitals which are oriented in some specific way.  For instance, liquid crystals are defined as such by the existence of orientational order, and resonant soft X-ray scattering studies using polarized light yield insight as to their orientational patterns.  In the realm of solids, non-cubic crystals generally have dichroism/birefringence whose axes depend on crystal axes.  This effect has been used to create contrast in X-ray images depending on crystal orientation.  Such images reveal important clues to the functionality of biomaterials such as nacre
and tooth enamel \cite{ma2009grinding,devol2014oxygen,stifler2018x,sun2017spherulitic}, or liquid crystals \cite{cao2002lasing}. 

An obvious next step is to combine the orientation sensitivity of resonant X-ray microscopy with the high spatial resolution of ptychography.  However, this program is not trivial \cite{ferrand2015ptychography,ferrand2018quantitative,baroni2019joint}.  Consider, for instance a material which is transparent but birefringent and has alternating stripes with the slow axis horizontal and vertical.
Now let the probe beam be polarized at 45$^\circ$ to the stripes.  The transmitted electric field consists of two components, one polarized along the same direction as the incident, and one polarized perpendicular ("depolarized").
The polarized component is uniform (assuming uniform thickness) but the depolarized component is uniform in intensity but has a pi phase shift between stripes.  Therefore, the far-field pattern is a sum of that from
a uniform material (spot at the $q=0$) and one from a phase grating (spots at multiples of $q=\lambda/period$).

There have been approaches to the reconstruction in which the measurement is repeated, with varying input polarizations, plus output polarization analyzers \cite{ferrand2015ptychography,ferrand2018quantitative} .  Such approaches work well in the visible-light and infra-red regime, but fail in the X-ray due to the lack of analyzers which work over a range of wavelengths.  What's needed is a reconstruction method which takes as input the set of diffraction patterns, taken at each pixel and at several input polarizations, and reconstructs the transmission matrix (Jones matrix) of the sample, without polarization analysis of the transmitted field.  In this paper, we propose such an algorithm and apply it to a simulated experiment on a realistic model of a biomineral sample.  We restrict ourselves to the case of linear dichroism only.  Ptychographic reconstruction of pure circular dichroism has been done, but it's easily shown that when the incident beam is circularly polarized and the sample only has circular dichroism, then there is no depolarization and the imaging is exactly as if the sample were isotropic. In this paper we will show how the vectorial optical properties of the sample may be reconstructed from a series of ptychographic scans with no analyzer, then specialize to the case of a single, uniaxially-anisotropic material.

\section{Dichroic ptychography}


In an isotropic medium with no optical anisotropy, the propagation of light through a thin sample may be described in terms of the electric field
\begin{equation}
\vec{E}(z)=\exp (ik  z n) \vec{E}_{0}
\label{eq:1}
\end{equation}
where  the beam is considered to propagate along $\widehat z$, $\vec E_0$ is the electric field vector at $z=0$, and $k=2 \pi / \lambda$, $\lambda, n$ have their usual meanings of wavelength and (complex) refractive index.  For an anisotropic medium, $n$ becomes a symmetric tensor, with eigenvectors defining the optical axes of the material.  There has been much work on the propagation of light through such media, starting with Maxwell's equations, for instance Berreman's $4\times 4$ matrix formulation 
\cite{berreman1972optics}.  This formulation is somewhat complex because it allows for optical activity and magnetic anisotropy.

One can simplify the equations considerably with the approximation that the index of refraction is very close to 1, and also that optical activity and magnetic effects are negligible, all of which are generally true  in non-magnetic systems.  In that case, sending a beam through a slab of material, even one with interfaces, does not produce a strong reflected wave.  Maxwell's equations then reduce to
\begin{equation}
\nabla \times \nabla \times \vec{E}=k^{2} \vec{D}=k^{2} \boldsymbol{\varepsilon} \vec{E}
\label{eq:2}
\end{equation}
where $\vec D$ is the electric displacement, $k=2\pi/\lambda$ , and $\epsilon$ is the dielectric tensor. We can remove the trivial z-dependence of the electric field by expressing it as
\begin{equation}
\vec{E}=\exp (i k z) \vec{F}(z)
\label{eq:3}
\end{equation}
We also decompose $\vec F$ into transverse and longitudinal parts $\vec{F}=\vec{F}_{T}+\hat{z} F_{Z}$ where $\vec F_T$  is perpendicular to the propagation direction ${\hat z}$.  Substituting into (2) yields the longitudinal component
\begin{equation}
\vec{F}_{Z}=\frac{i}{2} k \boldsymbol{\delta}_{T T} \vec{F}_{T}
\label{eq:4}
\end{equation}
where $\boldsymbol{\delta}=\boldsymbol{\varepsilon}-1$ and the subscript TT refers to the transverse components.  The equation for $\vec F_T$ is rather more complex:
\begin{equation}
-2 i k \vec{F}-\vec{F}^{\prime \prime}=k^{2} \boldsymbol{\delta}_{T T}+k^{2} \boldsymbol{\delta}_{T z} F_{Z}
\end{equation}
For a uniform medium, we can substitute in \eqref{eq:4} and solve the resulting eigenvalue problem.  However, under the approximation that all components of $\delta$ are much less than 1. we can drop all the complicating terms, ending up with
\begin{equation}
\vec{F}_{T}^{\prime}=\frac{i}{2} k \boldsymbol{\delta}_{T T} \vec{F}_{T}
\end{equation}
which has the solution
\begin{equation}
\vec{F}_{T}(z)=\exp \left[\frac{i k z}{2} \boldsymbol{\delta}_{T T}\right] \vec{F}_{T}(0)
\label{eq:7}
\end{equation}
which may be viewed as a tensor generalization of  Eq. \eqref{eq:1}.  Because the longitudinal component doesn't propagate outside the medium, we need only consider the transverse component when calculating the relation between the field incident on a flat sample and the field at the opposite surface.
	Considering a uniaxial material, the dielectric tensor may be expressed as
\begin{equation}
\begin{array}{c}{\varepsilon=\mathbf{n}^{2}} \\ {\mathbf{n}=n_{\perp}\mathbf{I}+\left(n_{ \|}-n_{\perp}\right) \widehat{m}^{T} \widehat{m}}\end{array}
\end{equation}
where $\widehat m$ is the direction of the optic axis of the crystal, $\mathbf I$ is the identity tensor, $\mathbf n$ is the refractive index tensor,  $\varepsilon$ is the dielectric tensor (the square of the refractive index tensor),  $n_\parallel, n_\perp$ are the indexes along and perpendicular to
the optic axis respectively.  For a uniaxial
material, two eigenvalues of $\mathbf n$ are equal ($n_{\perp_1}=n_{\perp_2}$) and the third one $n_\parallel$ is different from the first two.   For an isotropic material, the relation between the transmitted field and the incident is a scalar transmission.  
For a uniaxial system, in a coordinate system in which
the optic axis $\hat{m}$ is along $\widehat z$, it reduces to $\mathrm{diag}(n_{\perp_1},n_{\perp_2},n_{\parallel}).$

\begin{figure}
\begin{center}
\includegraphics[width=.95\textwidth]{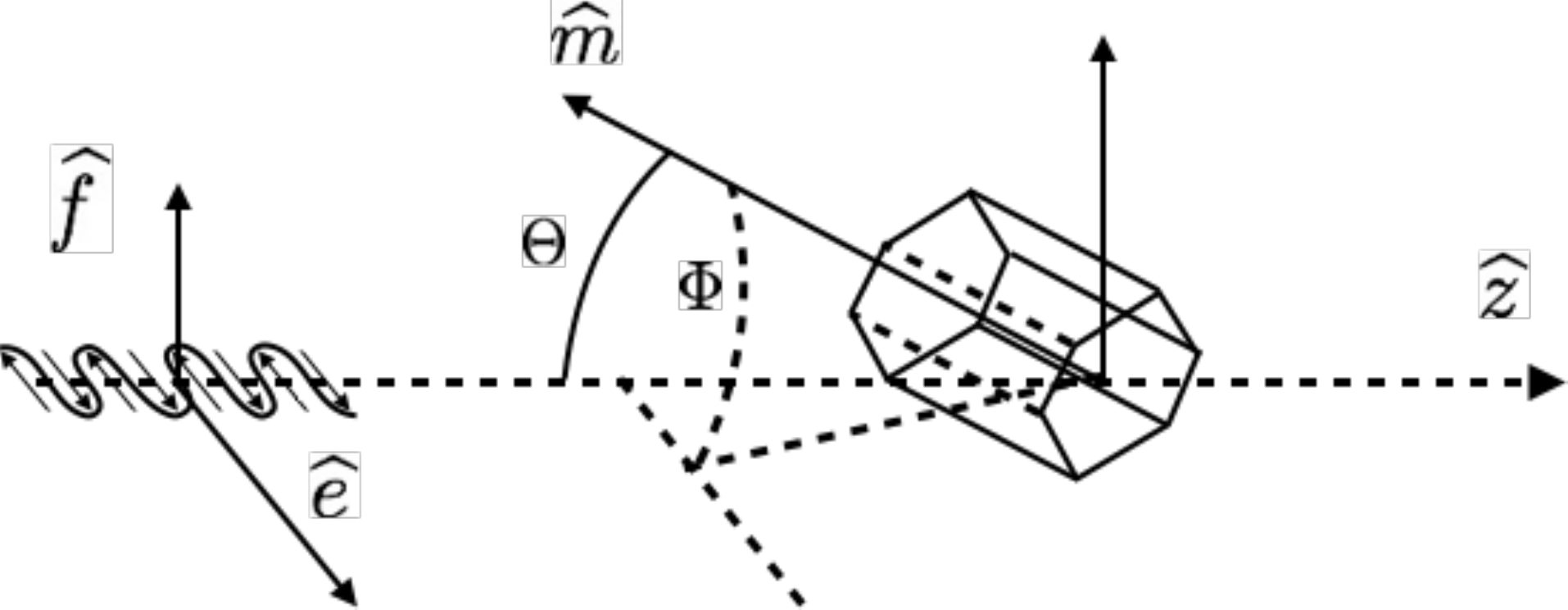}
\end{center}
\caption{Relationships between beam axis $\widehat{z}$, polarization direction $\widehat{e}$, and optical axis of the crystal $\widehat{m}$, and relative orientation angles ($\Theta$,$\Phi$).
}\label{fig0}
\end{figure}

 In the case of an anisotropic material, the transmission is a matrix which can be expressed in terms of polarized and depolarized parts.  If the beam propagates along the horizontal direction $\hat z$, and the incident field is along the horizontal direction $\hat e$ perpendicular to $\hat z$, $\vec{E}(0)={E_{0}} \hat{e}$, where $E_0$ is the electromagnetic field strength 
 then by calculating the matrix exponent in \eqref{eq:7} and inserting into \eqref{eq:3}, 
 the transmitted field is in general
\begin{equation}\label{eq:9}
\vec{E}(z=t)=\left(\hat{e} T_{1}+\hat{f} T_{2}\right) E_{0}
\end{equation}
where $t$ is the sample thickness and $\hat f=\widehat z \times \widehat e$ is a unit vector perpendicular to both $\hat e$ and the propagation direction $\widehat z$. The variables $T_{1}, T_2$ depend on the polarization angle $\chi$,  $\Theta$, and $\Phi$ 
are the polar and azimuth angles of the optical axis $\widehat m$ with respect to the 
beam direction $\hat z$ and polarization axis $\hat e$, shown in Fig \ref{fig0}:
\onecolumn
\begin{equation}
\begin{cases}
T_{1}(\chi,\Theta,\Phi)=\exp(\mathbf i k t n_{\perp} )(\cos^2(\Phi-\chi)\exp(\mathbf i k t\sin^2\Theta (n_\|-n_{\perp}))+\sin^2(\Phi-\chi));\\
T_{2}(\chi,\Theta,\Phi)=\tfrac12\exp(\mathbf i k t n_{\perp} )\sin2(\Phi-\chi)(\exp(\mathbf i k t\sin^2\Theta (n_\|-n_{\perp}))-1),
\end{cases}
\label{eq:t1t2}
\end{equation}
\twocolumn
Here $\Theta\in [0,\pi/2)$ and $\Phi\in[0, \pi)$ denote the polar angle of the axis relative to the beam, and the azimuthal angle of the crystal axis, respectively, which cannot  directly be measured. See Fig. \ref{fig0}.


It is the aim of ptychography to reconstruct what that transmission is for a set of points on the sample surface.  In dichroic ptychogrpahy, at a sequence of polarization angles $\chi_l\in [0, 2\pi), l=0, 1,\cdots, L-1$, the polarized and depolarized transmitted fields  
 ($T_{1,l}, T_{2,l}$) with $T_{1,l}:=T_{1}(\chi_l,\Theta,\Phi)$ are given by \eqref{eq:t1t2}.
The total intensity is $|\vec E|^2$ which is the sum of the magnitudes of the two perpendicular
polarizations. Moreover, the two polarizations add incoherently and do not interfere.
Hence {further by \eqref{eq:9}}, the dichroic ptychography measurements $I_{l}$ are given below, following the notations in \cite{chang2019blind}:
\begin{equation}
I_{l}=|\mathcal A(\omega, T_{1,l})|^2+|\mathcal A(\omega, T_{2,l})|^2,
\label{dich-ptycho}
\end{equation}
where
$\omega$ is the illumination, 
$\mathcal A(\omega, T_{1, l}):=(\mathcal A^T_0(\omega, T_{1, l}),$ $\mathcal A^T_1(\omega, T_{1, l})$, $\cdots$, $\mathcal A^T_{J-1}(\omega, T_{1, l}))^T,$ and $\mathcal A_j(\omega, T_{1,l}):=\mathcal F (\omega \circ \mathcal S_j T_{1, l}),$
with $\mathcal S_j$ being the scanning functions taking a small window from the whole sample.   In this paper, we will investigate how to determine $\Theta$  and $\Phi$ from the measurements $\{I_l\}$.


\section{A two-step reconstruction method}

Based on \eqref{dich-ptycho} and \eqref{eq:t1t2},   one can consider the following nonlinear optimization problem to determine $\Theta$ and $\Phi$ as:

\begin{minipage}{.4\textwidth}
\begin{equation}
\begin{split}
&\min_{\Theta,\Phi} \sum_{l=0}^{L-1}\tfrac12 \left\|\sqrt{I_{l}}-\sqrt{|\mathcal A(\omega, T_{1,l})|^2+|\mathcal A(\omega, T_{2,l})|^2}\right\|^2,\\
&s.t.\\
&\!\!\!\! \!\!T_{1,l}=\exp(\mathbf i k t n_{\perp} )(\cos^2(\Phi-\chi_l)\exp(\mathbf i k t\sin^2\Theta (n_\|-n_{\perp}))\\
&\hskip  6cm +\sin^2(\Phi-\chi_l)),\\
&\!\!\!\! \!\! T_{2,l}=\tfrac12\exp(\mathbf i k t n_{\perp} )\sin2(\Phi-\chi_l)(\exp(\mathbf i k t\sin^2\Theta (n_\|-n_{\perp}))-1),\\
&\forall~ 0\leq l\leq L-1.
\end{split} 
\label{eq:couple}
\end{equation}
\end{minipage}
\twocolumn

The above problem is highly nonlinear and nonconvex w.r.t. the variables $\Theta$ and $\Phi$, and 
it seems challenging to design  corresponding reliable algorithm. However, if considering the problem in a two-step way, one can 
first solve the two-mode ptychography problem without constraint to determine $T_{1,l}$ and $T_{2,l}$, and then determine $\Theta$ and $\Phi$ by solving these two constraints.


In the following,  a two step method is proposed below, which includes the two-modes ptychography reconstruction  to get $T_{1,l}$ and $T_{2,l}$ and then  determine the angles $\Theta$ and $\Phi$. 

We assume that these parameters including the $k$, $n_\perp$, $n_\parallel$ and the thickness map $t$ are known in advance.  We further remark  that the thickness map needs to be determined as follows. If considering the case $n_\|$  and $n_{\perp}$ are almost the same, i.e. $n_\|-n_{\perp}\approx 0,$  readily one has $T_1\approx\exp(\mathbf i ktn_\perp)$ and $T_2\approx 0$ by \eqref{eq:t1t2}. Hence by solving a standard ptychography reconstruction problem, one can derive $T_1$ and further get the thickness map $t$ with known $k, n_\perp$.

\subsection{Determining $T_{1,l}$ and $T_{2,l}$: two-mode ptychography}
In order to distinguish between $T_1$ and $T_2$ we 
 force $T_2=0$ and $T_1=1$ when the sample is empty. 

We use a modified ADMM \cite{chang2019blind} to solve the following two-mode ptychography problem in \eqref{eq:couple} (without constraints) as
\begin{equation}\label{eq:min}
\min_{\{T_{1,l}, T_{2,l}\}_{l=0}^{L-1}}  \sum_l\left\|\sqrt{I_l}-\sqrt{|\mathcal A(\omega, T_{1,l})|^2+|\mathcal A(\omega, T_{2,l})|^2} \right\|^2.  
\end{equation}
By introducing the auxiliary variable $z_{1,l}=\mathcal A(\omega, T_{1,l})$ and
$z_{2,l}=\mathcal A(\omega, T_{2,l})$, one gets the following equivalent form:
\begin{equation}
\begin{split}
&\min_{\{T_{1,l}, T_{2,l}\}_{l=0}^{L-1}}  \sum_l\left\|\sqrt{I_l}-\sqrt{|z_{1,l}|^2+|z_{2,l}|^2} \right\|^2,\\
&\qquad\qquad s.t.~ z_{1,l}=\mathcal A(\omega, T_{1,l}), z_{2,l}=\mathcal A(\omega, T_{2,l}),
\end{split}
\end{equation}
where $z_{1,l}$ and $z_{2,l}$ are auxiliary variables.
Then the augmented Lagrangian with multipliers $\Lambda_1$ and $\Lambda_2$ and penalization parameter $r>0$, can be given:
\begin{equation}
\begin{split}
&\mathscr L_r(\omega, T_{1,l}, T_{2,l}, z_{1,l}, z_{2,l};\Lambda_1,\Lambda_2):=
\sum_l\left\|\sqrt{I_l}-\sqrt{|z_{1,l}|^2+|z_{2,l}|^2} \right\|^2\\
&\qquad+r\sum_l\left( \Re\langle z_{1,l}-\mathcal A(\omega, T_{1,l}), \Lambda_{1,l}\rangle+ \Re\langle z_{2,l}-\mathcal A(\omega, T_{2,l}), \Lambda_{2,l}\rangle\right)\\
&\qquad+\tfrac{r}{2}\sum_l\left( \| z_{1,l}-\mathcal A(\omega, T_{1,l})\|^2+ \| z_{2,l}-\mathcal A(\omega, T_{2,l})\|^2\right),
\end{split}
\end{equation}
where $r$ is a positive constant, $\langle\cdot,\cdot\rangle$ denotes the inner product in complex Euclidean space, and $\Re$ denotes the real part of a complex-valued vector.
We can solve the following the saddle point problem instead of \eqref{eq:min} as 
\begin{equation}
\max_{\Lambda_{1,l}, \Lambda_{2,l}} \min_{\omega, T_{1,l}, T_{2,l}, z_{1,l}, z_{2,l}}   
\mathscr L_r(\cdot).
\end{equation}
Alternating minimizing the objective functions w.r.t. the variables $\omega$, $T_{1,l}$, $T_{2,l}$, $z_{1,l}$, $z_{2,l}$
and updating the multipliers yield the below two-mode ADMM algorithm:
\onecolumn
\begin{itemize}
\item [0.] Initialize the probe to be the average of all the frames of the intensities, $n:=0,$ and set $T^0_{1,l}:=\mathbf 1,$ and $T^0_{2,l}$ random.
$\gamma_1$ and $\gamma_2$ are two positive parameters. Set $\Lambda^n_{1,l}:=0,\Lambda^n_{2,l}:=0$ $\forall 0\leq l\leq L-1.$
\item [1.] Update illumination by
\begin{equation}
\omega^{n+1}=\frac{\gamma_1 \omega^n+ \sum_{l,j}\mathcal S_j (T^n_{1,l})^* \mathcal F^*(z^n_{1,l,j}+\Lambda^n_{1,l,j})+\mathcal S_j (T^n_{1,2})^* \mathcal F^*(z^n_{2,l,j}+\Lambda^n_{2,l,j}) }{\gamma_1\mathbf 1+\sum_{l,j} |\mathcal S_j T^n_{1,l}|^2+  |\mathcal S_j T^n_{2,l}|^2  }    
\end{equation}
\item[2.] Update $T_{1,l}$ and $T_{2,l}$ by
\begin{equation}
\begin{split}
 &T_{1,l}^{n+1}=  \frac{\gamma_2 T_{1,l}^n+\sum_{j}\mathcal S_j^T((\omega^{n+1})^*\mathcal F^*(z^n_{1,l,j}+\Lambda^n_{1,l,j}))}{\gamma_2\mathbf 1+\sum_j\mathcal S_j^T|\omega^{n+1}|^2};\\
 &T_{2,l}^{n+1}=  \frac{\gamma_2 T_{2,l}^n+\sum_{j}\mathcal S_j^T((\omega^{n+1})^*\mathcal F^*(z^n_{2,l,j}+\Lambda^n_{2,l,j}))}{\gamma_2\mathbf 1+\sum_j\mathcal S_j^T|\omega^{n+1}|^2}
\end{split}
\end{equation}

\item[3.] Update $z_1, z_2$ as 
\begin{equation}
\begin{split}
&z_{1,l}^{n+1}=\frac{\sqrt{I_l}+r\rho^{n}}{1+r}\circ \frac{\mathcal A(\omega^{n+1}, T^{n+1}_{1,l})-\Lambda^n_{1,l}}{ \rho_l^{n+1}};\\
&z_{2,l}^{n+1}=\frac{\sqrt{I_l}+r\rho^{n}}{1+r}\circ \frac{\mathcal A(\omega^{n+1}, T^{n+1}_{2,l})-\Lambda^n_{2,l}}{ \rho_l^{n+1}}
\end{split}
\end{equation}
where
$\rho_l^{n}:=\sqrt{|\mathcal A(\omega^{n+1}, T^{n+1}_{1,l})-\Lambda^n_{1,l}|^2+|\mathcal A(\omega^{n+1}, T^{n+1}_{2,l})-\Lambda^n_{2,l}|^2}$

\item[4.] Update multipliers as
\begin{equation}
\begin{split}
&\Lambda_{1,l}^{n+1}=\Lambda^n_{1,l}+r(z^n_{1,l}-\mathcal A(\omega^{n+1}, T_{1,l}^{n+1}));\\
&\Lambda_{1,2}^{n+1}=\Lambda^n_{2,l}+r(z^n_{2,l}-\mathcal A(\omega^{n+1}, T_{2,l}^{n+1})).
\end{split}    
\end{equation}

\item[5.] If  the stopping conditions are satisfied, then output $T_{1,l}^{n+1}, T_{2,l}^{n+1}$ as the final solutions; otherwise, set $n:=n+1$ and go to Step 1.

\end{itemize}
    \hrulefill

\twocolumn

To further remove ambiguities of $T_{1,l}$ (the phase factors) for different polarization angles, we assume that the sample has the empty region, where the contrast of $T_{1,l}$  keep as constant  and  that of $T_{2,l}$ as zeros. Therefore, additional projections are enforced in Step 2, where in the empty region, the iterative solution of $T_{1,l}$ sets to one , while $T_{2,l}$ sets to zeros.

\subsection{Solving $\Theta$ and $\Phi$}
In the following three subsections, we will show how to get $\Theta$ and $\Phi$ step by step.

By further reformulating these two constraints by introducing auxiliary variables $u, v, u_1, u_2$, one gets
\begin{equation}
\begin{cases}
T_{1,l}=u_1 \cos(2\chi_l)+u_2\sin(2\chi_l)+v;\\
T_{2,l}=-u_1\sin(2\chi_l)+u_2\cos(2\chi_l);\\
u_1=u\circ\cos(2\Phi);\\
u_2=u\circ \sin(2\Phi);
\end{cases}
\label{eq:reform}
\end{equation}
with $\circ$ denotes the Hadamard product (pointwise multiplication)
where
\begin{equation}
\label{eq:44}
\begin{split}
&u:=\tfrac12\exp(\mathbf i k t n_{\perp} )\circ(\exp(\mathbf i k t\sin^2\Theta (n_\|-n_{\perp}))-1);\\
&v:=\tfrac12\exp(\mathbf i k t n_{\perp} )\circ(\exp(\mathbf i k t\sin^2\Theta (n_\|-n_{\perp}))+1).\\
\end{split}
\end{equation}

With the prior of the empty region, the depolarized part $T_{2,l}$ still has phase ambiguity and we can only use the information of $T_{1,l}$ to recover the components of $u_1, u_2$ and $v$. 
By solving the following least squares problem
$\min_{u_1, u_2,v}\sum_l \|T_{1,l}-u_1 \cos(2\chi_l)-u_2\sin(2\chi_l)-v\|^2$,
one can readily get
\begin{equation}
E^TE
\begin{pmatrix}
u_1[m]\\
u_2[m]\\
v[m]
\end{pmatrix}
=E^T
\begin{pmatrix}
T_{1,0}[m]\\
\vdots\\
T_{1,L-1}[m]
\end{pmatrix}
\end{equation}
with 
\begin{equation}
E:=
\begin{pmatrix}
\cos(2\chi_0)&\sin(2\chi_0)&1\\
\vdots&\vdots&\vdots\\
\cos(2\chi_{L-1})&\sin(2\chi_{L-1})&1\\
\end{pmatrix}.    
\end{equation}
Finally one derives the 
\begin{equation}
\begin{pmatrix}
u_1[m]\\
u_2[m]\\
v[m]
\end{pmatrix}
=(E^TE)^{-1}E^T
\begin{pmatrix}
T_{1,0}[m]\\
\vdots\\
T_{1,L-1}[m]
\end{pmatrix}.
\end{equation}


Then we consider to determine $u$ by solving  \eqref{eq:reform}.
Readily one knows that $u=v-\exp(\mathbf i k t n_{\perp}).$ 
Finally, by \eqref{eq:reform} and \eqref{eq:44}, we have
\begin{equation}
\begin{split}
&\Phi=\mathrm{arctan}(\tan(\Phi))=\arctan(\tfrac{\sin(2\Phi)}{1+\cos(2\Phi)})\\
&\quad =\arctan(\tfrac{u\sin(2\Phi)}{u+u\cos(2\Phi)})=\arctan(\tfrac{u_2}{u+u_1});\\
&\Theta=\arcsin(\sqrt{|\log((u+v)\circ\exp(-\mathbf i k t n_{\perp} ))/\mathbf i kt(n_\|-n_{\perp})|}).\\
\end{split}
\end{equation}

%

\section{Simulations}

\begin{figure}
\begin{center}
\begin{tabular}{cc}
\includegraphics[width=.32\textwidth]{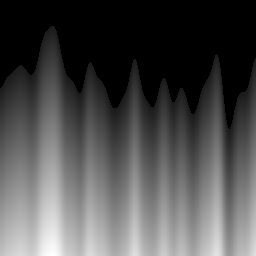} & 
\includegraphics[width=.32\textwidth]{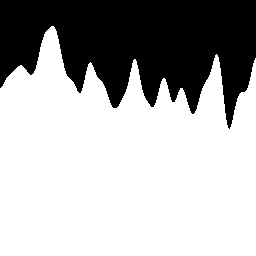}\\
(a) thickness & (b) mask
\end{tabular}
\begin{tabular}{ccc}
\includegraphics[width=.32\textwidth]{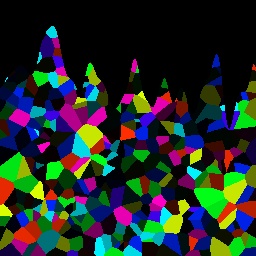} &
\includegraphics[width=.32\textwidth]{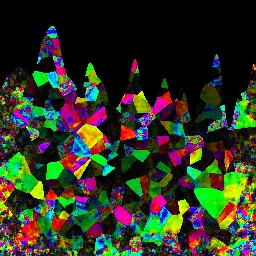} &
\includegraphics[width=.32\textwidth]{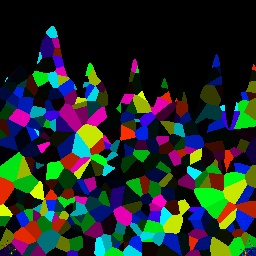}\\
(c) truth & (d) Scalar-Ptycho & (e) Tensor-Ptycho
\end{tabular}
\begin{tabular}{cc}
{\includegraphics[width=.1\textwidth]{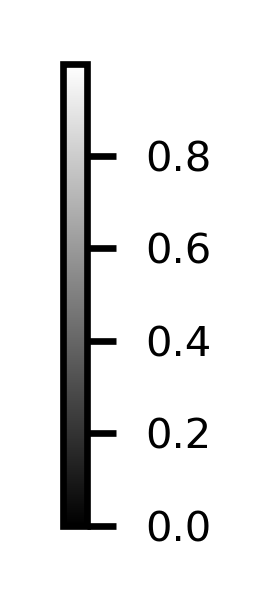}}
{\includegraphics[width=.32\textwidth]{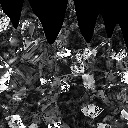}} &
{\includegraphics[width=.1\textwidth]{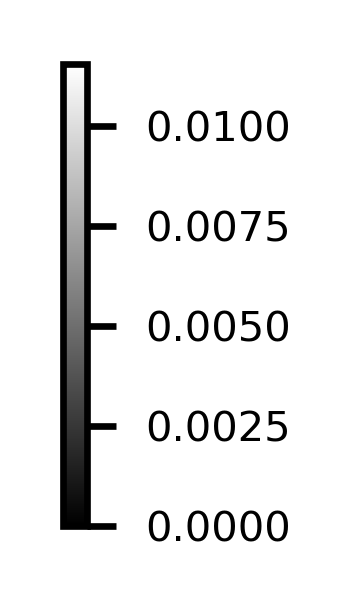}}
{\includegraphics[width=.32\textwidth]{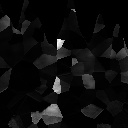}}\\
(f) err$_{map}$-Scalar & (g) err$_{map}$-Tensor
\end{tabular} 
\end{center}
\caption{Thickness map in (a) and mask or empty region in (b); Truth in (c) used for simulations with colors representing orientation \cite{devol2014oxygen}, and reconstructed results by scalar-ptychography in (d) and tensor-ptychography in (e), and error$_{map}$ for scalar-ptychography in (f) and (g) tensor-ptychography. Note the change of scale in reconstruction errors}
\label{fig1}
\end{figure}

Here we consider three polarization angles from three different angles $\chi_l\in\{0, \pi/3, 2\pi/3\}.$ 
The sample is based on a simulation of a FIB section of coral (random polycrystalline aragonite \cite{sun2017spherulitic}) with thickness gradient and curtaining ($256\times 256$ pixels shown in Fig. \ref{fig1}(c)). The thickness map is put in Fig. \ref{fig1}(a). We also use the mask for the empty region shown in Fig. \ref{fig1}(b). We used optical constants $n$ at the oxygen edge from the same paper, using \cite{watts2014calculation} to obtain the real part of the refractive index. For ptychography, we use the zone plate ($64\times 64$ pixels) with Half Width Half Max (HWHM)=18 pixels with raster scan with stepsize 16 pixels.

In order to measure the differences between the reconstructed angles $(\Phi_{rec}, \Theta_{rec})$ and the truth $(\Phi, \Theta)$,   we calculate the error map as:
\[
\mathrm{err}_{map}:=h\times h_{rec},
\]
with  
$h_{rec}:=(\sin\Theta_{rec}\circ\cos\Phi_{rec},\sin\Theta_{rec}\circ\sin\Phi_{rec}, \cos\Phi_{rec})$
and $h:=(\sin\Theta\circ\cos\Phi,\sin\Theta\circ\sin\Phi, \cos\Phi)$.
We also calculate the error maps for  $\Phi$ or $\Theta$, denoting by
$
\mathrm{err}_{\Phi}:=|\Phi-\Phi_{rec}|
$
and
$
\mathrm{err}_{\Theta}:=|\Theta-\Theta_{rec}|.
$

In order to show the recovered angles $(\Phi_{rec}, \Theta_{rec})$ in a color image, we use the synthetic HSV (Hue, Saturation, Value) format, where H and V channels are from $\Phi_{rec}$ and $\Theta_{rec}$ respectively.   

\graphicspath{{images/}}
\noindent {\bf {Noiseless test}}

We compare our proposed algorithm compared with the case named as ``scalar-ptycho'' when solving the scalar ptychography for different angles, neglecting the depolarized term, and then performing analysis as \cite{scholl2000observation}.

\begin{figure}
\begin{center}
\begin{tabular}{cc}
{\includegraphics[width=.1\textwidth]{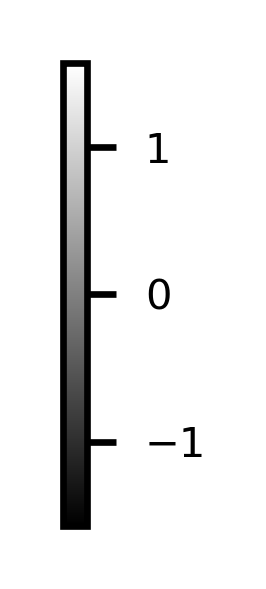}}
{\includegraphics[width=.35\textwidth]{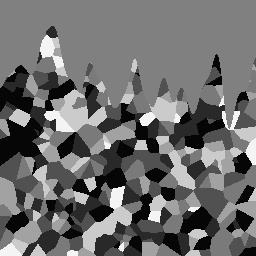}} &
{\includegraphics[width=.35\textwidth]{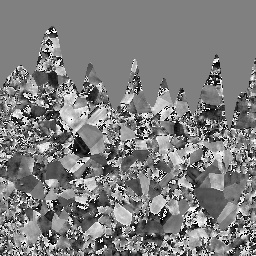}}\\
(a) $\Phi$ & (b) $\Phi_{rec}$-scalar\\
{\includegraphics[width=.1\textwidth]{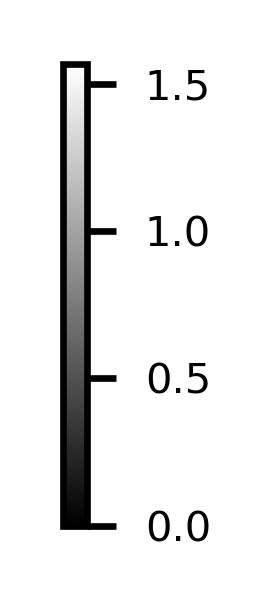}}
{\includegraphics[width=.35\textwidth]{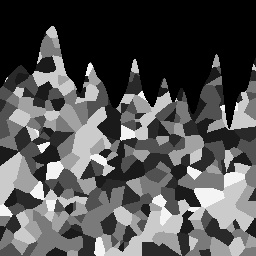}} &
{\includegraphics[width=.35\textwidth]{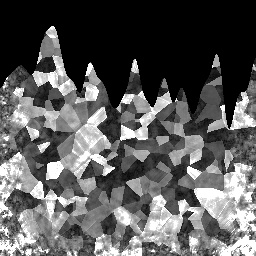}}\\
(c) $\Theta$ & (d) $\Theta_{rec}$-scalar \\
\includegraphics[width=.1\textwidth]{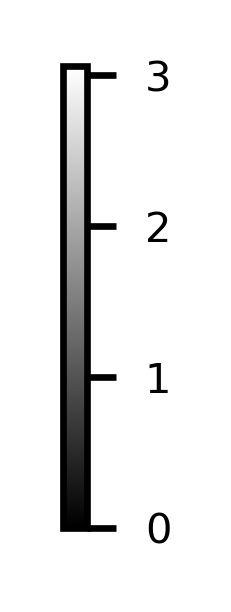}
{\includegraphics[width=.35\textwidth]{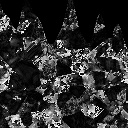}} &
{\includegraphics[width=.1\textwidth]{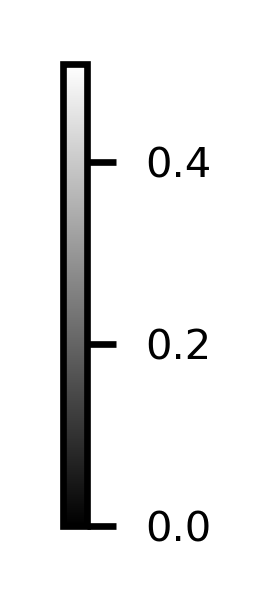}}
{\includegraphics[width=.35\textwidth]{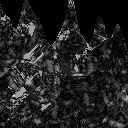}} \\
(e) $err_{\Phi}$ & (f) $err_{\Theta}$
\end{tabular}
\end{center}
\caption{Noiseless case for scalar-ptychography: comparison between ground truth used in simulations (left), and reconstruction using scalar ptychography (right). From top to bottom, (a,b) show the angle $\Phi$, (c,d) the angle $\Theta$, (e,f) the residual error of $\Phi,\Theta$ respectively}
\label{fig2}
\end{figure}

\begin{figure}
\begin{center}
\begin{tabular}{cc}
{\includegraphics[width=.1\textwidth]{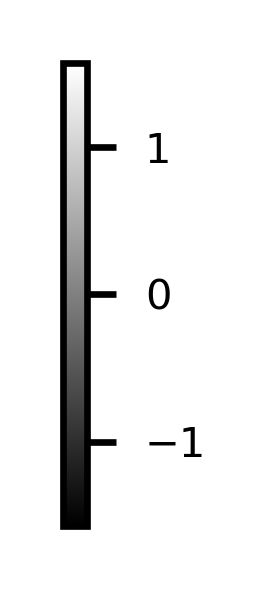}}
{\includegraphics[width=.35\textwidth]{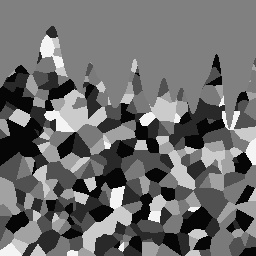}} &
{\includegraphics[width=.35\textwidth]{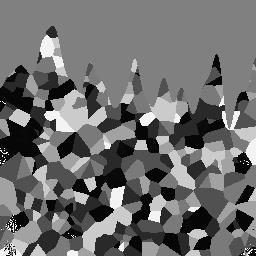}}\\
(a) $\Phi$ & (b) $\Phi_{rec}$ \\
{\includegraphics[width=.1\textwidth]{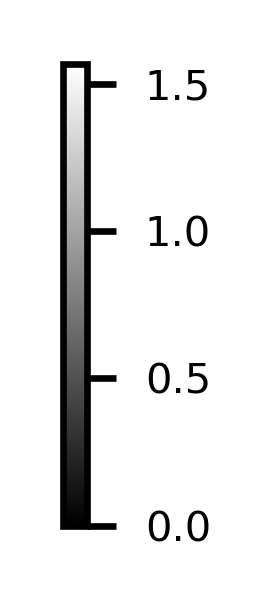}}
{\includegraphics[width=.35\textwidth]{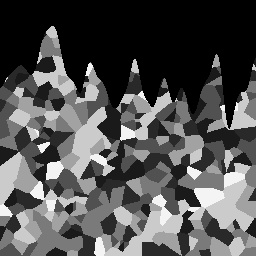}} &
{\includegraphics[width=.35\textwidth]{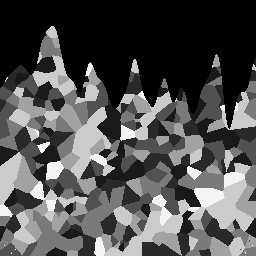}}\\
(c) $\Theta$ & (d) $\Theta_{rec}$ \\
{\includegraphics[width=.1\textwidth]{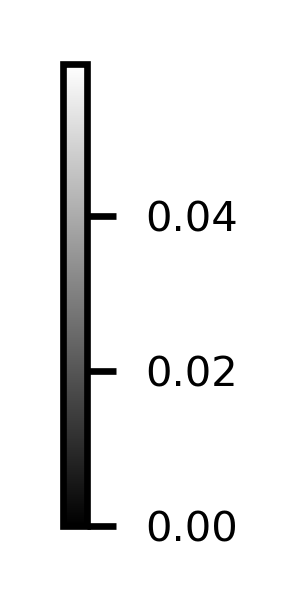}}
{\includegraphics[width=.35\textwidth]{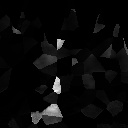}} &
{\includegraphics[width=.1\textwidth]{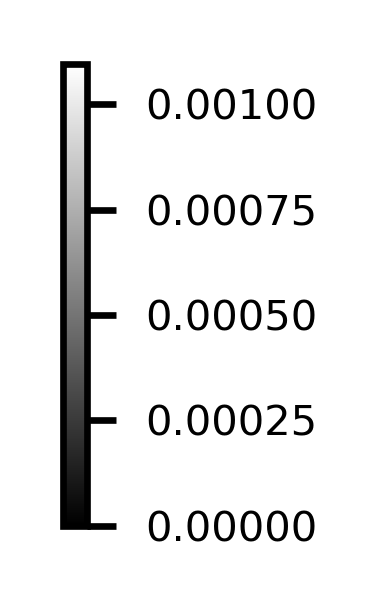}}
{\includegraphics[width=.35\textwidth]{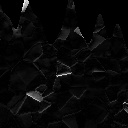}}\\
(e) $err_{\Phi}$ & (f) $err_{\Theta}$
\end{tabular}
\end{center}
\caption{Noiseless case for tensor-ptychography: comparison between ground truth used in simulations (left), and reconstruction using scalar ptychography (right). From top to bottom, (a,b) show the angle $\Phi$, (c,d) the angle $\Theta$, (e,f) the residual error of $\Phi,\Theta$ respectively}
\label{fig3}
\end{figure}

See Fig. \ref{fig1} for the reconstructed results by color images (HSV), and see more details for reconstructed angles in  Fig. \ref{fig2} by assuming scalar dielectric, and Fig. \ref{fig3} for the reconstruction using tensor ptychography. One can readily see that the proposed algorithm can give almost exact reconstruction compared with the scalar-ptychography.

\noindent {\bf {Noisy test}}

\begin{figure}
\begin{center}
\begin{tabular}{ccc}
{\includegraphics[width=.3\textwidth]{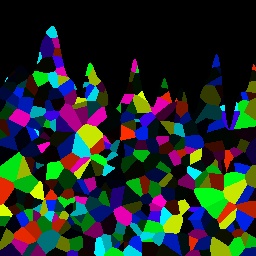}} &
{\includegraphics[width=.3\textwidth]{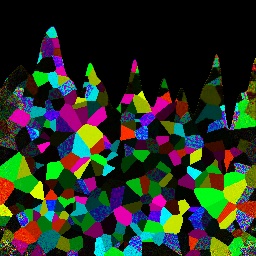}} &
{\includegraphics[width=.3\textwidth]{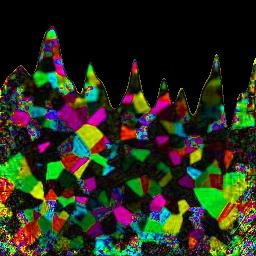}}\\
(a) Truth & (b) Reconstr & (c) Reconstr-Noisy \\
\includegraphics[width=.1\textwidth]{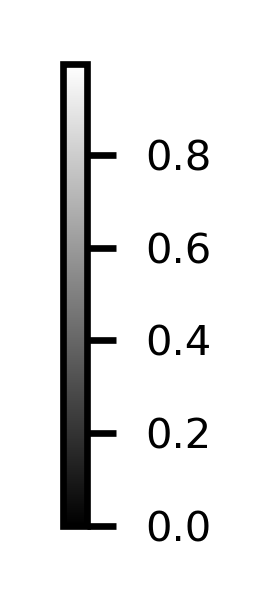}
&{\includegraphics[width=.3\textwidth]{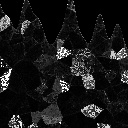}} &
{\includegraphics[width=.3\textwidth]{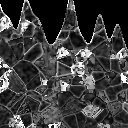}}\\
 & (e) err$_{map}$ & (f) err$_{map} -Noisy$
\end{tabular}
\end{center}
\caption{Noisy data: Truth in (a) and reconstructed results in (b) without noise, (c) from data contaminated by Poisson noise. (d,e) shows the error between ground truth and reconstruction for noiseless and noisy data}
\label{fig4}
\end{figure}

\begin{figure}
\begin{center}
\begin{tabular}{cc}
{\includegraphics[width=.1\textwidth]{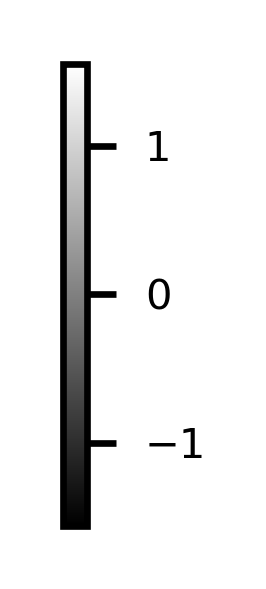}}
{\includegraphics[width=.35\textwidth]{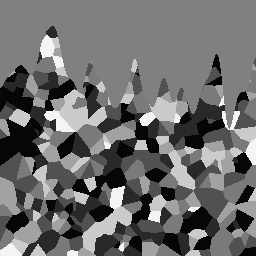}} &
{\includegraphics[width=.35\textwidth]{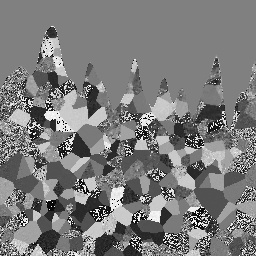}}\\
(a) $\Phi$ & (b) $\Phi_{rec}$ \\
{\includegraphics[width=.1\textwidth]{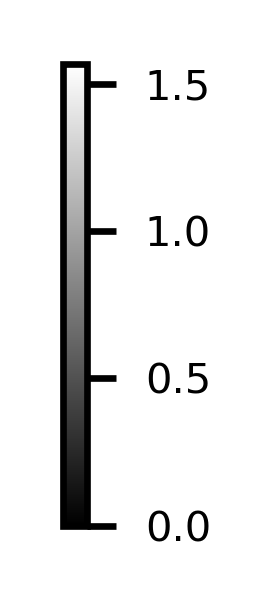}}
{\includegraphics[width=.35\textwidth]{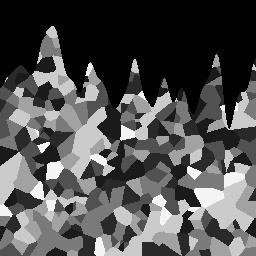}} &
{\includegraphics[width=.35\textwidth]{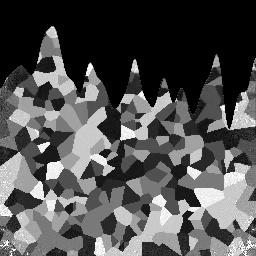}}\\
(c) $\Theta$ & (d) $\Theta_{rec}$ \\
{\includegraphics[width=.1\textwidth]{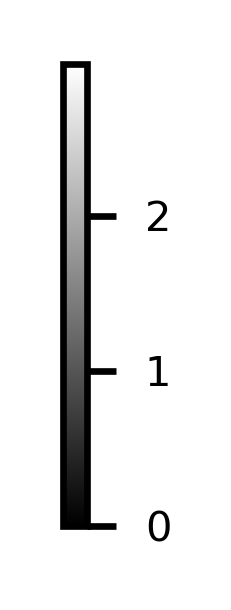}}
{\includegraphics[width=.35\textwidth]{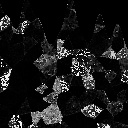}} &
{\includegraphics[width=.1\textwidth]{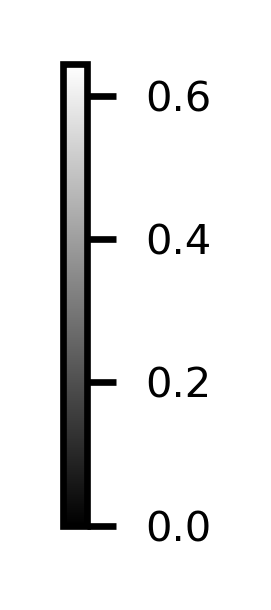}}
{\includegraphics[width=.35\textwidth]{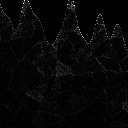}} \\
(e) $err_{\Phi}$ & (f) $err_{\Theta}$
\end{tabular}
\end{center}
\caption{Reconstructed angles from noisy data (SNR=48.2dB): (a,c) ground truth, (b,d) reconstruction from tensor ptychography, and (e,f) errors between truth and reconstruction}
\label{fig5}
\end{figure}

\begin{figure}
\begin{center}
\begin{tabular}{cc}
{\includegraphics[width=.1\textwidth]{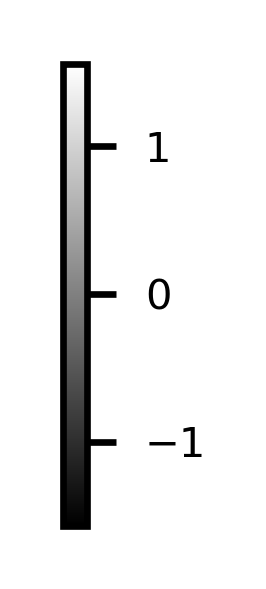}}
{\includegraphics[width=.35\textwidth]{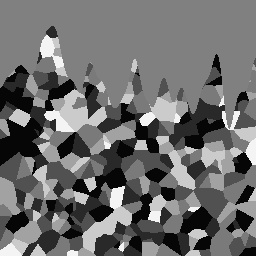}} &
{\includegraphics[width=.35\textwidth]{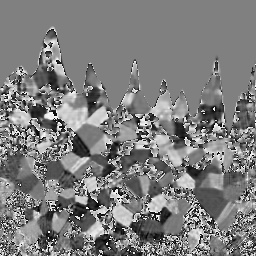}}\\
(a) $\Phi$ & (b) $\Phi_{rec}$ \\
{\includegraphics[width=.1\textwidth]{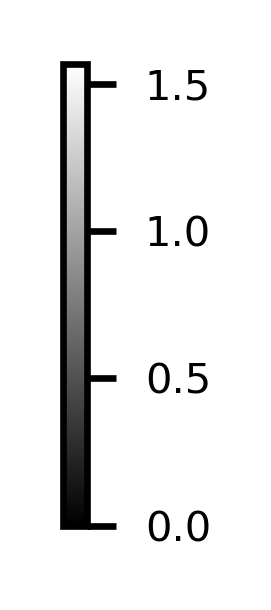}}
{\includegraphics[width=.35\textwidth]{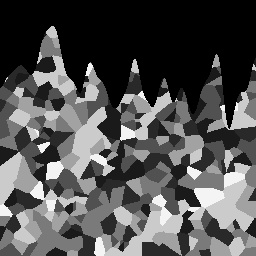}} &
{\includegraphics[width=.35\textwidth]{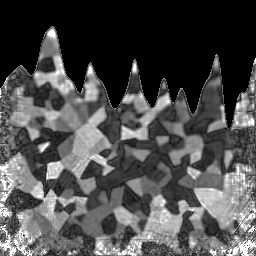}}\\
(c) $\Theta$ & (d) $\Theta_{rec}$ \\
{\includegraphics[width=.1\textwidth]{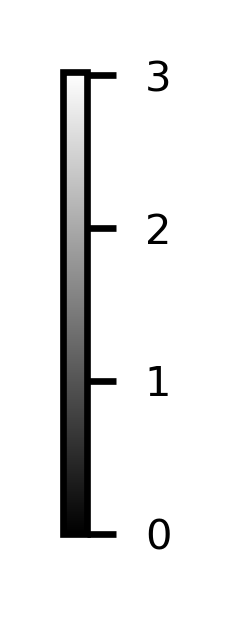}}
{\includegraphics[width=.35\textwidth]{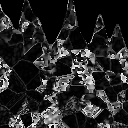}} &
{\includegraphics[width=.1\textwidth]{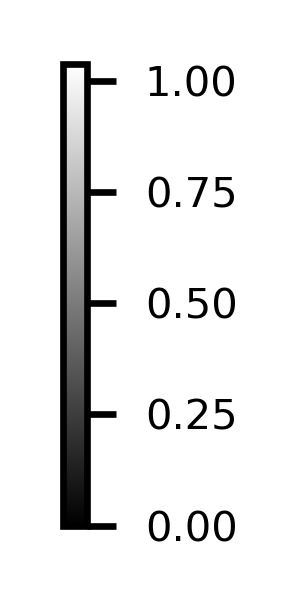}}
{\includegraphics[width=.35\textwidth]{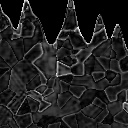}}
\\
(e) $err_{\Phi}$ & (f) $err_{\Theta}$
\end{tabular}
\end{center}
\caption{Reconstructed angles from noisy data (SNR=30.8dB): (a,c) ground truth, (b,d) reconstruction from tensor ptychography, and (e,f) errors between truth and reconstruction}
\label{fig6}
\end{figure}

We consider the data contaminated by different levels of Poisson noise,
the noise level are characterized by the signal-to-noise ratio (SNR) in dB, which is denoted below:
 \begin{equation}
 \mathrm{SNR}(X, X_c)=-10\log_{10}{\| X-X_g\|^2}/{\|X\|^2},
 \end{equation}
where $X$, $X_c$ correspond to the noisy and clean measurements respectively.
The results are shown in the color images in Fig. \ref{fig4}. The recovered angles are shown in Figs. \ref{fig5}-\ref{fig6} for different noise levels. One can readily see that as the noise levels increase, the recovery angles get less accurate and noisy.

\section{Conclusions}

We have shown that we are able to reconstruct the two components (polarized and depolarized) of the dielectric field at high resolution using ptychography and with the help of an empty region to fix the global phase factor ambiguity.
In order to obtain the reconstruction we have used a two step approach: first solve for different polarization independently, determine the orientations of $\Phi$ and $\Theta$ (within $\pm$ ambiguity) using the variables representing the dielectric constant obtained by least squares problems. 

The present algorithm assumes thickness  is known for each pixel and that the dielectric constants are known in advance.  In the future we may test methods for solving for thickness, determine optical constants such as measuring off resonance before and after the edge, using Kramers-Kronig relationships and  reference systems where the thickness and orientations are known in advance. The presented results used a two-step approach where we first reconstruct many polarized and depolarized transmissions for several polarizations. In the future we may directly solve \eqref{eq:couple} in order to enforce the consistency  between reconstructions which are determined by fewer parameters per pixels, such as shown in other multi-dimensional ptychography analysis such as tomography and spectroscopy. 
Furture work could also include taking data on a real sample and test reconstruction, adding denoising algorithms such as \cite{chang2018denoising,chang2016Total} and
developping tensor ptycho-tomography \cite{liebi2015nanostructure}, i.e. the reconstruction of the index of refraction tensor in 3D.

\section{Acknowledgments}

The simulated sample is based on an actual FIB section of coral provided by P. Gilbert who provided an estimated thickness map. 
B. Enders developed the software to generate a random distribution of patches with random orientations. 
H. Chang acknowledges the support of grant NSFC 11871372.
S. Marchesini acknowledges CAMERA  DE-AC02-05CH11231, US Department of Energy under Contract No. DE-AC02-05CH11231. The operations of the Advanced Light Source are supported by the Director, Office of Science, Office of Basic Energy Sciences, US Department of Energy under Contract No. DE-AC02-05CH11231. H. Chang would like to acknowledge the support from his host Professor James Sethian during the visit in LBNL.

\ifx \aff \undefined
\bibliographystyle{unsrt}
\bibliography{rD}
\else
\bibliographystyle{iucr}
\bibliography{rD}

@article{ferrand2018quantitative,
  title={Quantitative imaging of anisotropic material properties with vectorial ptychography},
  author={Ferrand, Patrick and Baroni, Arthur and Allain, Marc and Chamard, Virginie},
  journal={Optics letters},
  volume={43},
  number={4},
  pages={763--766},
  year={2018},
  publisher={Optical Society of America}
}

@article{ferrand2015ptychography,
  title={Ptychography in anisotropic media},
  author={Ferrand, Patrick and Allain, Marc and Chamard, Virginie},
  journal={Optics letters},
  volume={40},
  number={22},
  pages={5144--5147},
  year={2015},
  publisher={Optical Society of America}
}

@article{baroni2019joint,
  title={Joint estimation of object and probes in vectorial ptychography},
  author={Baroni, Arthur and Allain, Marc and Li, Peng and Chamard, Virginie and Ferrand, Patrick},
  journal={Optics Express},
  volume={27},
  number={6},
  pages={8143--8152},
  year={2019},
  publisher={Optical Society of America}
}

@article{sun2017spherulitic,
  title={Spherulitic growth of coral skeletons and synthetic aragonite: nature’s three-dimensional printing},
  author={Sun, Chang-Yu and Marcus, Matthew A and Frazier, Matthew J and Giuffre, Anthony J and Mass, Tali and Gilbert, Pupa UPA},
  journal={ACS nano},
  volume={11},
  number={7},
  pages={6612--6622},
  year={2017},
  publisher={ACS Publications}
}

@article{watts2014calculation,
  title={Calculation of the Kramers-Kronig transform of X-ray spectra by a piecewise Laurent polynomial method},
  author={Watts, Benjamin},
  journal={Optics Express},
  volume={22},
  number={19},
  pages={23628--23639},
  year={2014},
  publisher={Optical Society of America}
}

@article{scholl2000observation,
  title={Observation of antiferromagnetic domains in epitaxial thin films},
  author={Scholl, A and St{\"o}hr, J and L{\"u}ning, J and Seo, Jin Won and Fompeyrine, J and Siegwart, H and Locquet, J-P and Nolting, F and Anders, S and Fullerton, EE and others},
  journal={Science},
  volume={287},
  number={5455},
  pages={1014--1016},
  year={2000},
  publisher={American Association for the Advancement of Science}
}

@article{stifler2018x,
  title={X-ray Linear Dichroism in Apatite},
  author={Stifler, Cayla A and Wittig, Nina K{\o}lln and Sassi, Michel and Sun, Chang-Yu and Marcus, Matthew A and Birkedal, Henrik and Beniash, Elia and Rosso, Kevin M and Gilbert, Pupa UPA},
  journal={Journal of the American Chemical Society},
  volume={140},
  number={37},
  pages={11698--11704},
  year={2018},
  publisher={ACS Publications}
}

@article{cao2002lasing,
  title={Lasing in a three-dimensional photonic crystal of the liquid crystal blue phase II},
  author={Cao, Wenyi and Munoz, Antonio and Palffy-Muhoray, Peter and Taheri, Bahman},
  journal={Nature materials},
  volume={1},
  number={2},
  pages={111},
  year={2002},
  publisher={Nature Publishing Group}
}

@article{chang2016Total,
  title={Total Variation--Based Phase Retrieval for Poisson Noise Removal},
  author={Chang, Huibin and Lou, Yifei and Duan, Yuping and Marchesini, Stefano},
  journal={SIAM Journal on Imaging Sciences},
  volume={11},
  number={1},
  pages={24--55},
  year={2018},
  publisher={SIAM}
}

@article{ma2009grinding,
  title={The grinding tip of the sea urchin tooth exhibits exquisite control over calcite crystal orientation and Mg distribution},
  author={Ma, Yurong and Aichmayer, Barbara and Paris, Oskar and Fratzl, Peter and Meibom, Anders and Metzler, Rebecca A and Politi, Yael and Addadi, Lia and Gilbert, PUPA and Weiner, Steve},
  journal={Proceedings of the National Academy of Sciences},
  volume={106},
  number={15},
  pages={6048--6053},
  year={2009},
  publisher={National Acad Sciences}
}

@article{berreman1972optics,
  title={Optics in stratified and anisotropic media: 4$\times$ 4-matrix formulation},
  author={Berreman, Dwight W},
  journal={Josa},
  volume={62},
  number={4},
  pages={502--510},
  year={1972},
  publisher={Optical Society of America}
}

@article{devol2014oxygen,
  title={Oxygen spectroscopy and polarization-dependent imaging contrast (PIC)-mapping of calcium carbonate minerals and biominerals},
  author={DeVol, Ross T and Metzler, Rebecca A and Kabalah-Amitai, Lee and Pokroy, Boaz and Politi, Yael and Gal, Assaf and Addadi, Lia and Weiner, Steve and Fernandez-Martinez, Alejandro and Demichelis, Raffaella and others},
  journal={The Journal of Physical Chemistry B},
  volume={118},
  number={28},
  pages={8449--8457},
  year={2014},
  publisher={ACS Publications}
}

@article{liebi2015nanostructure,
  title={Nanostructure surveys of macroscopic specimens by small-angle scattering tensor tomography},
  author={Liebi, Marianne and Georgiadis, Marios and Menzel, Andreas and Schneider, Philipp and Kohlbrecher, Joachim and Bunk, Oliver and Guizar-Sicairos, Manuel},
  journal={Nature},
  volume={527},
  number={7578},
  pages={349},
  year={2015},
  publisher={Nature Publishing Group}
}

@article{rodenburg2008ptychography,
  title={Ptychography and related diffractive imaging methods},
  author={Rodenburg, John M},
  journal={Advances in imaging and electron physics},
  volume={150},
  pages={87--184},
  year={2008},
  publisher={Elsevier}
}

@article{chang2019blind,
  title={Blind Ptychographic Phase Retrieval via Convergent Alternating Direction Method of Multipliers},
  author={Chang, Huibin and Enfedaque, Pablo and Marchesini, Stefano},
  journal={SIAM Journal on Imaging Sciences},
  volume={12},
  number={1},
  pages={153--185},
  year={2019},
  publisher={SIAM}
}

@article{chang2018denoising,
  title={Denoising Poisson phaseless measurements via orthogonal dictionary learning},
  author={Chang, Huibin and Marchesini, Stefano},
  journal={Optics Express},
  volume={26},
  number={16},
  pages={19773--19796},
  year={2018},
  publisher={Optical Society of America}
}
\fi

\onecolumn

\twocolumn

\onecolumn

\twocolumn

\end{document}